\documentclass[sigconf]{acmart}
\acmConference[MSR 2024]{MSR '24: Proceedings of the 21st International Conference on Mining Software Repositories}{April 15–16, 2024}{Lisbon, Portugal}

\AtBeginDocument{%
 \providecommand\BibTeX{{%
   \normalfont B\kern-0.5em{\scshape i\kern-0.25em b}\kern-0.8em\TeX}}}
    
\copyrightyear{2024}
\acmYear{2024}
\setcopyright{acmcopyright}\acmConference[MSR 2024]{MSR '24: Proceedings of the 21st International Conference on Mining Software Repositories}{April 15–16, 2024}{Lisbon, Portugal}
\acmPrice{15.00}
\acmDOI{00.0000/0000000.0000000}

\usepackage{xspace}
\newcommand{\ie}{\textit{i.e., }}
\newcommand{\eg}{\textit{e.g., }}
\newcommand{\etal}{\textit{et al. \xspace}}
\usepackage{longtable}
\usepackage{subcaption}
\usepackage{alphalph}
\usepackage{amsmath,amsfonts}
\usepackage{algorithmic}
\usepackage{graphicx}
\usepackage{comment}
\usepackage{textcomp}
\usepackage{xcolor}
\usepackage{adjustbox}
\usepackage{textcomp}
\usepackage[numbered]{bookmark}
\usepackage{soul}
\usepackage{booktabs}
\usepackage{multirow}
\usepackage{float}
\usepackage[many]{tcolorbox}
\usepackage{fontawesome}
\usepackage{graphicx}
\usepackage{url}
\usepackage{tabularx} 
\usepackage[justification=centering]{caption}
\usepackage[switch]{lineno}
\usepackage{ragged2e}
\usepackage{dirtytalk}
\usepackage{multicol}
\usepackage{csquotes}
\usepackage{pgfplots, pgfplotstable}
\usepackage{pgf-pie}
\usepackage{colortbl}
\usepackage{xspace}
\usepackage{array}
\usepackage{balance}
\usepackage{supertabular} 
\newcolumntype{L}{>{\arraybackslash}m{16cm}}
\usepackage[flushleft]{threeparttable}
\usepackage{supertabular}
\usepackage{mdframed}
  \pgfplotstableread[row sep=\\,col sep=&]{
        interval & carT & sd\\
     }\mydata

\usetikzlibrary{shapes.geometric, arrows} 

\tikzstyle{startstop} = [rectangle, rounded corners, minimum width=3cm, minimum height=1cm,text centered, draw=black, fill=black!10]

\tikzstyle{io} = [trapezium, trapezium left angle=70, trapezium right angle=110, minimum width=3cm, minimum height=1cm, text centered,text width=2cm, draw=black]

\tikzstyle{process} = [rectangle, minimum width=3cm, minimum height=1cm, text centered, text width=3cm, draw=black]

\tikzstyle{decision} = [diamond, minimum width=3cm, minimum height=1cm, text centered,text width=2cm, draw=black]

\tikzstyle{arrow} = [thick,->,>=stealth]

\tikzstyle{database} = [cylinder, shape border rotate=90, draw=black,minimum height=2cm,minimum width=3cm, text centered,text width=0.6cm]

\newlength\q 

        {\begin{itemize}\setlength{\itemsep}{0pt}\setlength{\parsep}{0pt}
	\setlength{\topsep}{0pt}\setlength{\parskip}{0pt}}
	{\end{itemize}}
	{\begin{enumerate}\setlength{\itemsep}{0pt}\setlength{\parsep}{0pt}
	\setlength{\topsep}{0pt}\setlength{\parskip}{0pt}}
	{\end{enumerate}}
	{\begin{enumerate}\setlength{\itemsep}{0pt}\setlength{\parsep}{0pt}
	\setlength{\topsep}{0pt}\setlength{\parskip}{0pt}\setlength{\leftmargin}{0pt}}
	{\end{enumerate}}   

\newtcolorbox{boxK}{
    sharpish corners, 
    boxrule = 0pt,
    toprule = 4.5pt, 
    enhanced,
    fuzzy shadow = {0pt}{-2pt}{-0.5pt}{0.5pt}{black!35} 
}

\newcommand{\eman}[1]{\textcolor{violet}{{\it [Eman says: #1]}}}

\newcommand{\ali}[1]{\textcolor{green}{{\it [Ali says: #1]}}}

\author{Eman Abdullah AlOmar}
\affiliation{
    \institution{Stevens Institute of Technology}
    \city{Hoboken, New Jersey}
    \country{USA}
}
\email{ealomar@stevens.edu}

\author{Anushkrishna Venkatakrishnan }
\affiliation{
    \institution{Rochester Institute of Technology}
    \city{Rochester, New York}
    \country{USA}
}
\email{av3278@rit.edu}

\author{Mohamed Wiem Mkaouer}
\affiliation{
    \institution{University of Michigan-Flint}
    \city{Flint, Michigan}
    \country{USA}
}
\email{mmkaouer@umich.edu}

\author{Christian D. Newman}
\affiliation{
    \institution{Rochester Institute of Technology}
    \city{Rochester, New York}
    \country{USA}
}
\email{cnewman@se.rit.edu}

\author{Ali Ouni}
\affiliation{
    \institution{ETS Montreal, University of Quebec}
    \city{Montreal, Quebec}
    \country{Canada}
}
\email{ali.ouni@etsmtl.ca}

\renewcommand{\shortauthors}{AlOmar et al.}

\begin{document}

\title{How to Refactor this Code? An Exploratory Study on Developer-ChatGPT Refactoring Conversations}



\begin{abstract}
Large Language Models (LLMs), like ChatGPT, have gained widespread popularity and usage in various software engineering tasks, including refactoring, testing, code review, and program comprehension. Despite recent studies delving into refactoring documentation in commit messages, issues, and code review, little is known about how developers articulate their refactoring needs when interacting with ChatGPT. In this paper, our goal is to explore conversations between developers and ChatGPT related to refactoring to better understand how developers identify areas for improvement in code and how ChatGPT addresses developers' needs. Our approach relies on text mining refactoring-related conversations from 17,913 ChatGPT prompts and responses, and investigating developers' explicit refactoring intention. Our results reveal that (1) developer-ChatGPT conversations commonly involve generic and specific terms/phrases; (2) developers often make generic refactoring requests, while ChatGPT typically includes the refactoring intention; and (3) various learning settings when prompting ChatGPT in the context
of refactoring.  We envision that our findings contribute to a broader understanding of the collaboration between developers and AI models, in the context of code refactoring, with implications for model improvement, tool development, and best practices in software engineering. 
\end{abstract}



\maketitle

\section{Introduction}
\label{Section:Introduction}

 Artificial intelligence has been reshaping educational and industrial landscapes, and Large Language Models (LLMs) are emerging as the main driving force behind this revolution. The ability to harness massive amounts of information in multiple modalities allows LLMs to perform various tasks that would typically require human intervention \cite{fan2023large,zhao2023survey,hou2023large,nathalia2023artificial,kasneci2023chatgpt,ahmad2023towards,roy2023unveiling,treude2023she,al2023ab,hou2023large,palacio2023evaluating}.

\vspace{-.3cm}
\begin{figure}[th]
\centering 
\includegraphics[width=\columnwidth]{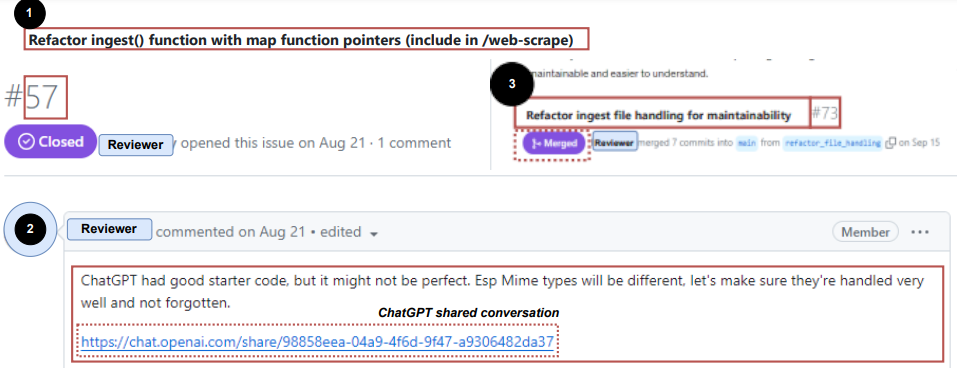}
\vspace{-.5cm}
\caption{Example of a ChatGPT conversation in the context of GitHub issue about refactoring \cite{Example}.} 
\label{fig:example}
\vspace{-.4cm}
\end{figure}
Learning from software repositories has opened up a myriad of LLM applications in software development tasks, such as code search \cite{wangayou}, code quality \cite{white2023prompt,white2023chatgpt,alam2023gptclonebench}, repair \cite{zhang2023critical,haque2023potential,sobania2023analysis,xia2023keep}, program comprehension \cite{ma2023scope},  generation \cite{feng2023investigating,dong2023self}, completion \cite{pudari2023copilot}, and translation \cite{jiao2023chatgpt}. The impressive performance of these LLMs, in general, and ChatGPT, in particular, has rapidly increased its popularity within the development community. According to a recent GitHub survey with 500 US-based developers \cite{futurism}, 
 92\% stated that their workflow has already integrated AI tools, and 70\% found these tools to increase their productivity. When asked about the reason behind this fast adoption, respondents said that they believe these AI tools will improve the quality of their code while meeting existing performance standards with fewer production-level incidents.

Although developers can assess the performance of generated code from a functional perspective \cite{feng2023investigating}, little is known about how the model would react to refactoring requests. By definition, refactoring is the practice of improving the internal code structure, without altering its external behavior \cite{Fowler:1999:RID:311424}. Given the subjective nature of refactoring, where there could be multiple equivalent solutions for a given input situation, it is interesting to see how LLMs would react to refactoring requests and what quality attributes they would target when optimizing the code.

The goal of this paper is to identify which of the various quality attributes, presented in the refactoring literature, are sought by developers when they request ChatGPT to refactor their code. Similarly, when developers request the refactoring of a given input code, we want to extract the quality attributes that ChatGPT reports optimizing when performing the refactoring. An illustrative example is shown in Figure \ref{fig:example}: An issue was opened to update the \texttt{ingest()} method. When querying ChatGPT, the model has provided an updated version of the code, while explicitly stating that the refactoring was intended to improve \textit{maintainability}. This can also be seen in the final title of the pull request that later integrated the modified code in production. Based on this example, we can see that \textit{maintainability} was one of the quality attributes that ChatGPT considers for code optimization, and so we want to extract all other quality attributes to better understand the refactoring strategies adopted by the model.

\vspace{-.2cm}
\section{Study Design}
\label{Section:Methodology}





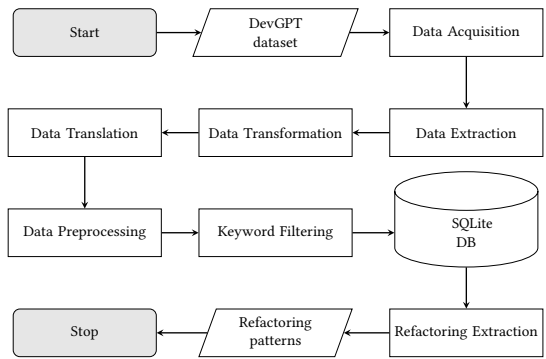
\begin{figure}[h]
\centering
\begin{adjustbox}{width=0.4\textwidth}
\begin{tikzpicture}[node distance=2cm]

\node (start) [startstop] {Start};
\node (io) [io, right of=start, xshift=2cm] {DevGPT dataset};
\node (pro1) [process, right of=io,xshift=2cm] {Data Acquisition};
\node (pro2) [process, below of=pro1, yshift=-0.1cm] {Data Extraction};
\node (pro3) [process, left of=pro2, xshift=-2cm] {Data Transformation};
\node (pro4) [process, left of=pro3, xshift=-2cm] {Data Translation};
\node (pro5) [process, below of=pro4, yshift=-0.1cm] {Data Preprocessing};
\node (pro6) [process, right of=pro5, xshift=2cm] {Keyword Filtering};
\node (db1) [database, right of=pro6,xshift=2cm] {SQLite DB};
\node (pro7) [process, below of=db1, yshift=-0.1cm] {Refactoring Extraction};
\node (io2) [io, left of=pro7, xshift=-2cm] {Refactoring patterns};
\node (stop) [startstop, left of=io2, xshift=-2cm] {Stop};

\draw [arrow] (start) -- (io);
\draw [arrow] (io) -- (pro1);
\draw [arrow] (pro1) -- (pro2);
\draw [arrow] (pro2) -- (pro3);
\draw [arrow] (pro3) -- (pro4); 
\draw [arrow] (pro4) -- (pro5);
\draw [arrow] (pro5) -- (pro6);
\draw [arrow] (pro6) -- (db1);
\draw [arrow] (db1) -- (pro7);
\draw [arrow] (pro7) -- (io2);
\draw [arrow] (io2) -- (stop);
\end{tikzpicture}
\end{adjustbox}
\vspace{-.2cm}
\caption{Overview of our experiment design.}
\label{fig:approach}
\vspace{-.4cm}
\end{figure}

Our study uses the DevGPT \cite{xiao2023devgpt} dataset
, which contains a wide range of information for open-source projects, such as code files, commits, issues, Hacker News, pull requests, and discussions. All, except Hacker News, utilize GitHub as their version control repository when demonstrating developer interactions with ChatGPT. Figure \ref{fig:approach} depicts a general overview of our experimental setup. In the following subsections, we elaborate on the activities involved in the process. 
\subsection{Dataset Construction}
To obtain data from various sources, we follow these steps:


\noindent\textbf{Step \#1: Data Acquisition:} The data were initially gathered from the \href{https://zenodo.org/records/10086809} {DevGPT dataset}. The dataset consists of multiple JSON files organized into snapshots.

\noindent \textbf{Step \#2: Data Extraction:} JSON files were extracted and consolidated into separate groups according to their source type.

\noindent\textbf{Step \#3: Data Transformation:} The JSON files were processed to transform them into a structured relational tabular form and loaded into the database. 

\noindent \textbf{Step \#4: Data Translation:} This step involves leveraging the deep
translator library's Google Translator\footnote{\url{https://deep-translator.readthedocs.io/en/latest/usage.html\#google-translate}} to facilitate the translation of non-English content into English. This multilingual analysis enhances the comprehensibility and accessibility of the dataset. Following the translation process, the database is systematically updated with the translated text.

\noindent\textbf{Step \#5: Data Preprocessing:} Titles, bodies, prompts, and responses from various source types were cleaned, stop-words removed, and tokenized.

To execute all these steps, we built a pipeline that takes as input the JSON files, and outputs the needed subset in the form of a database. Our pipeline uses various technologies, such as \textit{SQLite} for data management, \textit{FastText} for identifying non-English conversations, \textit{Google Translator} library for their translation, \textit{NLTK \& Spacy} for meticulous cleaning and tokenization, complemented by the efficiency of \textit{Dask} for concurrent processing in both cleaning and tokenization tasks. The pipeline and the generated database are available for replication and extension purposes\footnote{\url{https://smilevo.github.io/self-affirmed-refactoring/}}. 

\subsection{Refactoring Patterns Extraction }

To identify refactoring documentation patterns, we perform a series of manual and automated activities: 

\vspace{.1cm}
\noindent \textbf{Data sources associated with developer intention about refactoring.} As our study focuses on refactorings, our analysis is limited to source types where refactorings were discussed as part of developer-ChatGPT conversations. Hence, we first extracted all the different conversations from the source dataset. To ensure that the selected software  artifacts are about refactoring and to reduce the occurrence of false positives, we focused on prompts that reported developers' intention about the application of refactoring (\textit{i.e.}, having the keyword `\textit{refactor}'). The choice of `\textit{refactor}', besides being used by all related studies, is intuitively the first term to identify ideal refactoring-related commits \cite{murphy2008gathering,alomar2019can,di2018preliminary,Ratzinger:2008:RRS:1370750.1370759,alomar2021toward,alomar2021icse}. This step resulted in the selection of three source types: commits, issues, and files.  
The procedure led to the examination of 470 commits, 69 issues, and 176 files. 

\vspace{.1cm}
\noindent \textbf{Annotation of developer-ChatGPT conversations.} When using ChatGPT, developers use natural language to describe the software development artifacts. \textcolor{black}{Before running the experiment, a pilot experiment is conducted to ensure that the annotators reach a consensus on data annotation}. Hence, given the diverse nature of developers who describe the problem, an automated approach to analyzing the prompt and answer text is not feasible. Therefore, we performed a thematic analysis approach of the prompt and response to identify refactoring documentation patterns based
on guidelines provided by Cruzes \etal \cite{cruzes2011recommended}. Thematic analysis is
one of the most used methods in Software Engineering literature
(\eg \cite{Silva:2016:WWR:2950290.2950305,calefato2023lot}), which is a technique for identifying and recording patterns (or \say{themes}) within a collection of descriptive labels. Next, we grouped this subset of conversations based on specific patterns. 
 Further, to avoid redundancy of any pattern, we only considered one phrase if we found different patterns with the same meaning. For example, if we find patterns such as `simplifying the code', `code simplification', and `simplify code', we add only one of these similar phrases to the list of patterns. This enables us to have a list of the most insightful and unique patterns, and it also helps to create more concise patterns that are usable to the readers. Furthermore, we manually analyzed conversation patterns (\ie learning settings) between developers and ChatGPT when seeking guidance on refactoring tasks. Two authors independently annotated the conversations, and any conflicts in their annotations were subsequently resolved through discussion.

\vspace{-.1cm}
\section{Experimental Results}
\label{Section:Result}

\subsection{RQ$_1$: What textual patterns do developers use to describe their refactoring needs using ChatGPT?}
\noindent\textbf{Approach.} We manually inspect GitHub commits, issues, files to identify refactoring documentation patterns. These patterns are represented as a keyword or phrase frequently occurring in developer-ChatGPT conversations, as described in Section \ref{Section:Methodology}.


\begin{table}
\centering
\caption{List of refactoring documentation in DevGPT (`*' captures the extension of the keyword).}
\label{Table:GeneralPatterns}
\vspace{-.3cm}
\small
\begin{adjustbox}{width=0.5\textwidth,center}

\begin{tabular}{lllllll}
\toprule
\textbf{Patterns}\\
\midrule
Add* & Chang* & Chang* the name & Cleanup  & Clean* up \\
Code clarity & Code clean* & Code organization & Code review  &  Clean code \\
Creat*  & Customiz* & Easier to maintain & Encapsulat* & Enhanc*  \\
Extend* & Extract* & Fix* & Inlin* & Improv* \\
Improv* code quality & Introduc* & Merg* & Modif* & Modulariz* \\
 Migrat* & Mov* & Organiz* & Polish* & Reduc* \\
Refactor* & Refin* &  Remov* & Remov* redundant code & Renam*\\
Remov* unused dependencies  & Reorganiz* & Replac* & Restructur* & Rework* \\
Rewrit* & Simplif* & Split* \\

 \hline
       
\end{tabular}
\end{adjustbox}
\vspace{-.2cm}
\end{table} 

\vspace{-.3cm}
\begin{figure}[h]
\centering 
\includegraphics[width=0.6\columnwidth]{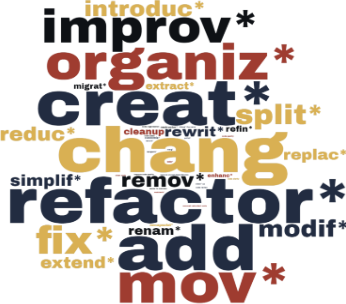}
\vspace{-.4cm}
\caption{Popular refactoring textual patterns.}
\label{fig:wordcloud}
\vspace{-.4cm}
\end{figure}

\noindent\textbf{Results.} Our in-depth inspection resulted in a list of 43 refactoring documentation patterns, as shown in Table~\ref{Table:GeneralPatterns} and Figure \ref{fig:wordcloud}. Our findings show that the names of refactoring operations (\textit{e.g.}, `extract*', `mov*', `renam*') occur in the top frequently occurring patterns, and these patterns are mainly linked to code elements at different levels of granularity such as classes, methods, and variables.  These specific terms are well-known software refactoring operations and indicate developers' knowledge of the catalog of refactoring operations. We also observe that the top-ranked refactoring operation-related keywords include `mov*', `renam*', and `extract*'. Moreover, we observe the occurrences of refactoring specific terms such as `cleanup', `code quality', and `restructur*'.   The observation from this research question aligns with the work
of Russo \cite{russo2023navigating}, who found that a significant portion of software professionals indicated their intention to use LLMs to improve and maintain their codebase.

\begin{boxK}
\textit{\textbf{Summary for RQ$_1$.} Developer-ChatGPT conversations involve diverse textual patterns about refactoring activities. These patterns encompass generic descriptions like `improve code readability' as well as specific refactoring operation names following Fowler's conventions, including `extract' and `rename'.}
\end{boxK}

\subsection{RQ$_2$: What quality attributes does ChatGPT consider when describing refactoring?}
\noindent\textbf{Approach.} After identifying the different refactoring documentation patterns, we categorize the patterns into three main categories (similar to previous studies \cite{alomar2019can,alomar2021ESWA}): (1) internal quality attributes, (2) external quality attributes, and (3) code smells. 
\begin{table}
\fontsize{6.8}{6.8}\selectfont
\begin{center}
\caption{Summary of refactoring patterns, clustered by refactoring related categories.}
\label{Table:TopSpecificKeywords}
\vspace{-.3cm}
\begin{tabular}{lll}

\toprule
\textbf{Internal QA (\%)} &
\textbf{External QA  (\%)} &
\textbf{Code Smell  (\%)} \\
         \midrule 
    
  Dependency (25.33\%)    & Readability (22.13\%) &  Code smell (91\%)  \\ 
   Inheritance (25\%)     &  Usability (16.19\%)    &   Long method (6\%) \\ 
   Composition (14.16\%)      &  Performance (10.25\%)  & Duplicate code (3\%) \\ 
    Abstraction (12.16\%)   & Maintainability (8.32\%)    &   \\ 
   Coupling (10\%)  & Flexibility (8.32\%)   &  \\
   Encapsulation (8.83\%)   &  Reusability (7.28\%)   &  \\
   Polymorphism (3\%) &  Accessibility (7.28\%)   & \\
   Complexity (1.5\%)  &  Modularity (4.6\%)   & \\
     & Extensibility (3.86\%)  &  \\
      & Correctness (1.78\%)  & \\ 
         & Manageability (1.48\%) & \\
             & Robustness (1.48\%)  & \\
     & Compatibility (1.33\%)   &  \\
    &   Scalability (1.18\%) & \\
    & Configurability (0.89\%)   &  \\
        & Simplicity (0.89\%) & \\
    & Reliability (0.89\%) & \\
    & Productivity (0.89\%) & \\
    & Adaptability (0.59\%)  & \\
    & Understandability (0.29\%) & \\

        \bottomrule
\end{tabular} 
\end{center}
\vspace{-.6cm}
\end{table}

\noindent\textbf{Results.} Table \ref{Table:TopSpecificKeywords} provides the list of refactoring documentation patterns, ranked based on their frequency, which we identify in ChatGPT responses. We observe that ChatGPT frequently mentions key internal quality attributes (such as `inheritance', `complexity', etc.), a wide range of external quality attributes (such as `readability' and `performance'), and code smells (such as `duplicate code' and `long method') that might impact code quality. To improve internal design, optimization of the structure of the system with respect to its dependency and inheritance appears to be the dominant focus that is consistently mentioned in the conversation (25.33\% and 25\%, respectively). Concerning external quality attribute, we observe the mention of refactorings to enhance nonfunctional attributes. Patterns such as `readability', `usability', and `performance' represent the ChatGPT' main focus, with 22.13\%, 16.19\%, and 10.25\%, respectively. The focus on code readability  from the ChatGPT may be attributed to the fact that the model frequently tends to
generate code that adheres to common naming and coding conventions, which was highlighted in a previous study when assessing the readability of the ChatGPT code \cite{dantas2023assessing}. Finally, for code smells, ChatGPT commonly used the generic term `code smell' with 91\%, the `long method', and `duplicate code' code smells represent the most popular anti-pattern ChatGPT intends to describe refactoring (6\% and 3\%, respectively). Furthermore, by analyzing the ChatGPT response, we notice that the model emphasizes certain coding practices, including dependency injection, naming conventions, exception handling, single responsibility principle, unit tests, separation of concerns, design patterns, and behavior preservation.
\begin{boxK}
\textit{\textbf{Summary for RQ$_2$.} Our findings indicate that developers frequently make generic requests when seeking guidance on refactoring tasks, while ChatGPT typically includes the intention when applying refactorings, addressing quality issues related to internal attributes, external attributes, or code smells, along with applying well-known coding practices.}
\end{boxK}

\vspace{-.3cm}
\subsection{RQ$_3$: How do developers typically initiate conversations with ChatGPT when seeking guidance on refactoring tasks?}
\noindent\textbf{Approach.} To explore whether specific conversation patterns emerge as developers collaborate with ChatGPT for iterative refinement of refactoring solutions, we manually examine refactoring conversations and categorize the identified patterns.

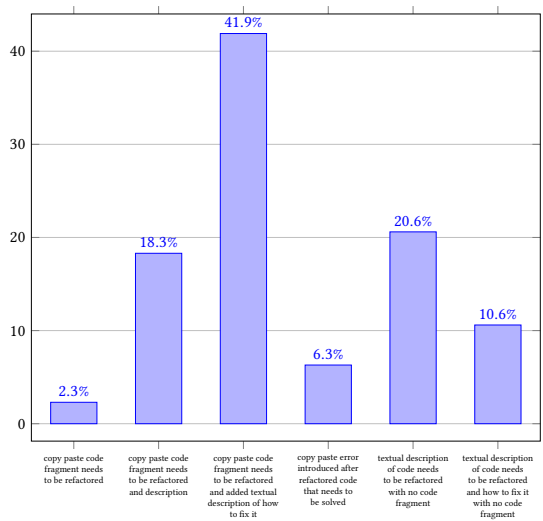
\begin{figure}
\centering
     \pgfplotstableread[row sep=\\,col sep=&]{
        interval & carT \\
        A & 2.3  \\
        B & 18.3  \\
        C & 41.9 \\
        D & 6.3  \\
        E & 20.6 \\
         F & 10.6 \\
     }\mydata

      \begin{tikzpicture}
      \begin{scope}[scale=0.70]
          \begin{axis}[
              ybar,
              ymajorgrids,
              bar width=25pt,
              width=320pt,
              ymax=44,
              symbolic x coords={A,B,C,D,E,F},
              xtick=data,
              xticklabel style={align=center,font=\tiny},
              xticklabels={{copy paste code \\ fragment needs \\ to be refactored},
                           {copy paste code \\ fragment needs \\ to be refactored \\ and description},
                           {copy paste code \\ fragment needs \\ to be refactored \\ and added textual \\ description of how \\ to fix it},
                           {copy paste error \\ introduced after \\ refactored code \\ that needs to \\ be solved},
                           {textual description \\ of code needs \\ to be refactored \\ with no code \\ fragment},
                           {textual description \\ of code needs \\ to be refactored \\ and how to fix it \\ with no code \\ fragment}},
              nodes near coords={\pgfmathprintnumber\pgfplotspointmeta \%},
              nodes near coords align={vertical},
              ]
              \addplot+[error bars/.cd, y dir=both, y explicit] table[x=interval,y=carT, y error=sd]{\mydata};
          \end{axis}
          \end{scope}
      \end{tikzpicture}
      \vspace{-.3cm}
      \caption{ChatGPT conversation patterns to refactor code.}
\label{fig:prompt pattern}
\vspace{-.3cm}
\end{figure}

\noindent\textbf{Results.} The effectiveness of any language model relies heavily on the ability of developers to craft appropriate prompts. Therefore, one of the outcomes of this work is to raise awareness among developers of the importance of prompt engineering, particularly in the context of refactoring. Analysis of developer prompts (as depicted in Figure \ref{fig:prompt pattern}) reveals that 2.3\% of developers simply copy and paste code fragments that need to be refactored, assuming that ChatGPT can discern the intention behind the code. 18.3\% of developers copy and paste code fragments that need to be refactored along with textual description/instruction to provide context. Most developers (41.9\%) copy and paste code fragments that need to be refactored and add a textual description of how to fix it. About 6.3\% of developers introduce copy and paste errors after the refactoring application suggested by ChatGPT. Some developers (20.6\%) add textual description of the code that needs to be refactored, but without adding the associated code fragments. Others (10.6\%) add a textual description of the code that needs to be refactored and how to fix it, but also without adding the associated code fragments.

\begin{boxK}
\textit{\textbf{Summary for RQ$_3$.} Many developers (41.9\%) tend to copy and paste code fragments that require refactoring, along with providing a textual description of how they want them to be addressed.}

\end{boxK}

\section{Discussion \& Conclusion}
\label{Section:Discussion}

\noindent\textbf{ Takeaway \#1:  \textit{ ChatGPT limited understanding of the broader context of the codebase.}} Although ChatGPT provides an informative context of refactorings, its comprehension of the entire codebase is restricted, leading to possible missing dependencies and codependencies. This limitation becomes apparent in instances where ChatGPT makes suggestions based on misunderstandings or false assumptions about the code. This exploratory study serves to unveil this practical limitation, encouraging developers to grasp the model's mechanics rather than treating it as a black box that consistently generates acceptable answers. The \textbf{RQ$_1$} and \textbf{RQ$_2$} results not only show instances where developers express their refactoring needs for certain queries, but also reveal potential issues with suggested code changes. Some developers encountered compiler errors introduced by ChatGPT-provided code, and in some cases, the suggested refactoring led to test failures, contrary to the expectation that refactoring should preserve the system's internal behavior.  

\noindent\textbf{ Takeaway \#2:  \textit{ The quality of ChatGPT responses is highly dependent on the quality of the prompts.}} The effectiveness of ChatGPT is closely related to the quality of its training dataset \cite{fan2023large}. According to the developer-ChatGPT conversations (\textbf{RQ$_1$} and \textbf{RQ$_2$}), ChatGPT offers insightful suggestions on refactorings, which is particularly helpful for addressing non-urgent issues like code style, adding comments, and more. On the other hand, challenges arise when ChatGPT misunderstands the reported code fragments that need to be refactored or becomes confused when receiving a code excerpt instead of the entire code, especially when the code is too large to provide. Its corrections sometimes also misunderstood the purpose of an excerpt and would slightly alter the logic. Therefore, the more complex the given code input, the more likely that the code output will not work properly. By identifying common patterns and challenges in these refactoring conversations, researchers and developers can work toward improving ChatGPT's capabilities (\eg following software documentation quality framework \cite{treude2020beyond}). This understanding can guide the refinement of the model to better address developers' needs.

\noindent\textbf{ Takeaway \#3:  \textit{ Variability in refactoring conversation learning settings between developers and ChatGPT.}} From \textbf{RQ$_3$},  we observe that some of the developers' prompts were zero-shot, where they relied on the model's generative ability to understand an issue or propose a corresponding code fix. Zero-shot learning involves the model making decisions on presumably unseen data by approximating it with previously trained code \cite{xian2017zero}. For instance, the prompt asks the model to refactor the code to eliminate duplication of an input source code. Further, the prompt can be augmented by adding a label to the unseen data, \ie one-shot \cite{fei2006one}. For example, mentioning how large file splitting can be performed can guide the model towards the decision to take (class extraction). Some developers opted for few-shot learning, entering code fragments and asking ChatGPT to propose fixes while checking for design antipatterns (\eg duplicated code, bad names, dead code) by providing an example. Few-shot learning involves providing context-specific examples to help the model make better decisions about complex tasks \cite{wang2020generalizing}. For instance, the input of code changes that address the complexity of a given class. This RQ shows the non-uniformity of developers' prompts, with a majority debating how to extract the necessary action from the models, while others overestimate the model's capabilities. This aligns with previous studies that have demonstrated the susceptibility of ChatGPT to hallucinations when it comes to coding semantic structures \cite{ma2023scope,white2023chatgpt,yang2023harnessing,hou2023large,kou2023model}. Insights from these conversations can inform the integration of ChatGPT or similar models into development tools, making them more user-friendly and aligned with developers' expectations during refactoring tasks.

\noindent\textbf{ Takeaway \#4: \textit{Need for better refactoring descriptions.}} 
 During our manual analysis, we observed various cases where developers ask the model to refactor a given input code by only mentioning the word refactoring (\eg refactor this...). While it allows for a high flexibility, it is better to constrain the model with what the developer expects. For instance, Adding the intent (why?) behind the refactoring, and potential instructions (how?) can help with ensuring specific coding practices and styles that are targeted by the developer's own decisions or their company policy.

\bibliographystyle{ACM-Reference-Format}
\bibliography{sample-base}

\end{document}